\documentclass[conference]{IEEEtran}%
\usepackage{amsfonts}
\usepackage{amsmath}
\usepackage{amssymb}
\usepackage{graphicx}
\usepackage{cite}%
\providecommand{\U}[1]{\protect\rule{.1in}{.1in}}
\newtheorem{theorem}{Theorem}

\newtheorem{definition}[theorem]{Definition}

\newtheorem{remark}[theorem]{Remark}

\begin{document}

\title{Competitive Privacy in the Smart Grid:\ An Information-theoretic Approach}
\pubid{~}
\specialpapernotice{~}

%

\author{\authorblockN
{Lalitha Sankar, Soummya Kar, Ravi Tandon, H. Vincent Poor}
\authorblockA{Dept. of Electrical Engineering,\\
Princeton University,
Princeton, NJ 08544.\\
{\{lalitha,skar,rtandon,poor\}}@princeton.edu\\}}%
%

\maketitle
%

\begin{abstract}%

\footnotetext{The research was supported in part by the Air Force Office of
Scientific Research under MURI Grant FA9550-09-1-0643, in part by the Army
Research Office under MURI Grant W911NF-11-1-0036, and in part by the National
Science Foundation under Grants CNS-09-05398 and CCF-10-16671.
\par
{}}Advances in sensing and communication capabilities as well as power
industry deregulation are driving the need for distributed state estimation at
the regional transmission organizations (RTOs). This leads to a new
\textit{competitive privacy} problem amongst the RTOs since there is a tension
between sharing data to ensure network reliability (utility/benefit to all
RTOs) and withholding data for profitability and privacy reasons. The
resulting tradeoff between utility, quantified via fidelity of its state
estimate at each RTO, and privacy, quantified via the leakage of the state of
one RTO\ at other RTOs, is captured precisely using a lossy source coding
problem formulation for a two RTO network. For a two-RTO model, it is shown
that the set of all feasible utility-privacy pairs can be achieved via a
single round of communication when each RTO communicates taking into account
the correlation between the measured data at both RTOs. The lossy source
coding problem and solution developed here is also of independent interest.%

\end{abstract}%

\section{Introduction}

\begin{figure}
[ptb]
\begin{center}
\includegraphics[
height=2.6057in,
width=3.224in
]%
{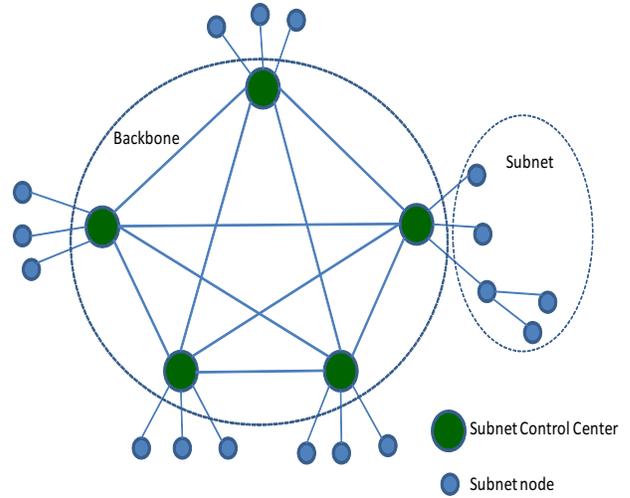}%
\caption{Mutiple RTOs (subnets) connected via the transmission network.}%
\label{FigNet}%
\end{center}
\end{figure}

The electric power industry is undergoing profound changes as greater emphasis
is placed on the importance of a smarter grid that supports sustainable energy
utilization. Technically, enabled by advances in sensing, communication, and
actuation, power system estimation and control are likely to involve many more
fast information gathering and processing devices (e.g. Phasor Measurement
Units) \cite{Bose2010SmartGrid}. Economically, the deregulation of the
electricity industry has led to the creation of many regional transmission
organizations (RTOs) within a large interconnected power system \cite{Wu2005}
(see Fig. \ref{FigNet}). Both technical and economic drivers suggest the need
for more distributed estimation and control in power system operations.%

\textit{Distributed Cooperation}: While the distributed state estimation
problem has been investigated for well over two decades, the focus to date has
been on two-tier hierarchical models \cite{cutsem1981ps} in which each local
control center (e.g., RTO) estimates independently without sharing any
information amongst its neighbors, and at a higher level, a central
coordinator receives the estimation results from the individual areas and
coordinates them to obtain a system-wide solution. However, such an
hierarchical approach does not scale with increasing measurement rates due to
communication and reliability challenges inherent in systems with one
coordination center. Furthermore, the interconnectedness of the RTOs makes the
problem of wide area monitoring and control (e.g. for the entire Eastern
United States interconnection) important and immediate. This requirement of
sharing the entire system state amongst all the RTOs is driving the need for a
fully distributed approach to state estimation wherein the local control
centers interactively estimate the system state as a whole. \ 

There are several challenges that arise in the context of a fully distributed
state estimation approach. We address a specific problem of collaboration and
competition by introducing a framework that precisely quantifies the tradeoff
between the utility (benefit) of collaboration and the resulting privacy leakage.

\textit{Competitive Privacy}: The problem of end-user privacy has begun
receiving attention with the deployment of smart meters to monitor and finely
manage user power consumption
\cite{Quinn:SSRN:09,SGPMag3,SGPProc1,SGPProc2,SGPArt1,TimeSeries2}. In
contrast, a new \textit{competitive privacy} problem arises at the level of
the RTOs due to the conflict between sharing data for distributed estimation
and withholding data for economic (competitive) and end-user privacy reasons,
i.e., there is a tradeoff between sharing data to ensure network reliability
(utility/benefit to all RTOs) and withholding data to ensure profitability and privacy.

Competitive privacy has been studied by the database community in the context
of data integration from multiple autonomous sources that do not wish to
reveal private sensitive business information to one another and yet benefit
from the combined analysis \cite{CompPrivacy}. However, the solutions proposed
are unique for databases and do not directly apply to the distributed state
estimation problem discussed above. An analytical framework that abstracts the
problem of competitive privacy via appropriate source models and utility and
privacy metrics is required and it is this problem that we address in this paper.

\textit{Utility vs. Privacy}: Utility and privacy are competing goals: utility
is maximized when the RTOs completely disclose their measurements to each
other; however, privacy is minimal for this disclosure. On the other hand, not
sharing any data guarantees maximal privacy but achieves zero utility. Thus,
ensuring privacy requires distorting or perturbing the data which also provide
utility guarantees. The theory of rate distortion allows us to study this
tradeoff between privacy and utility using appropriate metrics for both
constraints; we present such a formal framework in this paper for the
competitive privacy problem.

\textit{Contributions}: In this paper, we present a linear measurement model
for the grid at the level of the RTOs that takes into account the
interconnections amongst them. Viewing the power system state at each RTO as
an information source, we model the measurements at each RTO as a linear
combination of all the sources. The competitive privacy problem formulated
thus leads to a new \textit{distributed lossy source coding problem} wherein
each RTO shares (encodes) a perturbed function of its measurements, and over a
finite number of such communication rounds, estimates its state subject to two
constraints: a) each RTO must be able to decode its own measurements to a
desired fidelity, and b) at each RTO, the privacy leakage of the measurements
shared by the other RTOs must be bounded.

Distributed source coding has received much attention recently with a growing
interest in sensor networks. However, with a few exceptions, most distributed
source coding problems remain open. One such exception is the quadratic
Gaussian CEO\ problem in which a source is observed noisily by multiple
sensors, each of which uses a finite-rate link to communicate it to a single
receiver which in turn combines these messages to reconstruct the source to a
desired level of fidelity \cite{QGCEO}. The problem presented here further
generalizes such a setting in the following ways: a) there are multiple
sources, b) each sensor (here RTO) is now both an encoder and a decoder, i.e.,
there are multiple encoders and decoders, and c) there are fidelity
requirements at each RTO for its own measurements and privacy requirements
vis-\`{a}-vis other measurements.

Having modeled the general problem, we focus on a two-source setting that
captures the distributed estimation problem for two large\ RTOs. Using
distortion as a metric for utility and mutual information as a metric for
privacy leakage, we demonstrate that a single-shot rate-distortion code at
each RTO\ that is cognizant of the side information (measurement vector) at
the other RTO achieves the set of all feasible utility-privacy operating points.

The paper is organized as follows: in Section \ref{SecMM}, we develop the
model and present abstract metrics for utility and privacy. In Section
\ref{SecMR}, we present our main results. We conclude in Section \ref{SecCR}.

\section{\label{SecMM}Model and\ Metrics}

\subsection{Model}

Let $M$ denote the total number of RTOs and let $X_{m}$ denote the power
system state variable at the $m^{th}$ RTO, $m=1,2,\ldots,M$. Assuming
linearized system dynamics, the general measurement model at the $k^{th}$
terminal at a sampling time instant is given by
\begin{equation}
Y_{m}=%
{\textstyle\sum\limits_{j=1}^{M}}
H_{m,j}X_{j}+Z_{m},\text{ }m=1,2,\ldots,M, \label{mod_y}%
\end{equation}
where $X_{m}\sim\mathcal{N}\left(  0,1\right)  $, for all $m,$ are assumed to
be mutually independent, and where $H_{m,j}$ denotes the Jacobian modeling the
linearized dynamics between the $j^{th}$ state and the measurement at the
$m^{th}$ RTO. The observation noise at the $m^{th}$ RTO $Z_{m}\sim
\mathcal{N}\left(  0,\sigma_{m}^{2}\right)  $ is assumed to be independent of
$X_{m}$, for all $m$. We assume that the $H_{m,j}$ are fixed and known at all
RTOs for the duration of communications; furthermore, we also assume that the
statistics of the noise, the measurements, and the states are known at all the
RTOs. In this paper, we focus on case of $M=2$, i.e., we are interested in the
distributed estimation problem between two (typically large) adjacent RTOs.
Finally, we set $H_{1,1}=H_{2,2}=1$, $H_{1,2}=\alpha\in\left(  0,\infty
\right)  $, and $H_{2,1}=\beta\in\left(  0,\infty\right)  $.

\begin{figure}
[ptb]
\begin{center}
\includegraphics[
height=2.3618in,
width=2.7752in
]%
{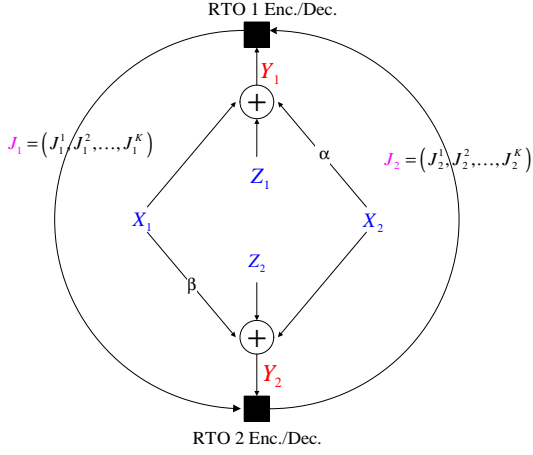}%
\caption{Communication protocol for two RTOs. }%
\label{FigTwoRTO}%
\end{center}
\end{figure}

We assume that the $m^{th}$ RTO observes a sequence of $n$ measurements
$Y_{m}^{n}=[Y_{m,1}$\textbf{ }$Y_{m,2}$ $\ldots$ $Y_{m,n}]$, for all $m$, over
a window of time prior to communications. The communication protocol is shown
in Fig. \ref{FigTwoRTO} for $M=2$. In each communication round, the two RTOs
take turns such that each RTO encodes (quantizes) its measurements taking into
account the fact the other RTO has correlated side information (the
correlations come from the measurement vectors at each RTO\ being functions of
the state variables at both RTOs). We assume a total of $K$ such communication
rounds before RTO $m$ estimates $\hat{X}_{m}^{n}$, $m=1,2$, using the entire
sequence of public communications and its own measurements.%

\subsection{Metrics}

\textit{Utility}: The motivation for cooperation at each RTO is to obtain a
good estimate of its own system state. Thus, for our continuous Gaussian
distributed state model, a reasonable metric for utility at the $m^{th}$ RTO
is the mean square error $D_{m}$ of the original and the estimated state
sequences $X_{m}^{n}$ and $\hat{X}_{m}^{n}$, respectively.

\textit{Privacy}: The measurements at each RTO in conjunction with the
quantized data from the other RTO while enabling estimation of its own state
can also potentially leak information about the state at the other RTO. We
capture this information leakage via the mutual information.

\subsection{Communication Protocol}

A coding scheme for the network involves an encoder-decoder pair at each RTO
over $K$ communication rounds such that some desired utility and privacy
values are achieved. Formally, an $(n,K,M_{1},M_{2},D_{1},D_{2},L_{1},L_{2})$
code for this network results from a $K$-round protocol in which RTO $m$,
$=1,2,$ uses the encoding functions $\left\{  f_{m,k}\right\}  $ for the
$k^{th}$ round and a decoding function $F_{m}$ at the end of the $K$ rounds.
In round $k$, the encoder at RTO $m$ maps its measurements and the messages
received until then from the other RTO to an index set $\mathcal{J}_{m}^{k}$
where
\begin{equation}
\mathcal{J}_{m}^{k}\equiv\left\{  1,2,\ldots,J_{m}^{k}\right\}  \text{,
}k=1,2,\ldots K,\text{ }m=1,2, \label{Enc1}%
\end{equation}
is the index set at the $m^{th}$ RTO for mapping the quantized sequence in the
$k^{th}$ protocol round via the encoder $f_{m}^{k}$ defined as%
\begin{equation}
f_{m}^{k}:\mathcal{Y}_{m}^{n}\times\mathcal{J}_{m}^{1}\times\ldots
\times\mathcal{J}_{m}^{k-1}\rightarrow\mathcal{J}_{m}^{k},\text{ }k=1,2,\ldots
K,m=1,2, \label{Enc2}%
\end{equation}
such that at the end of the $K$ rounds, the decoding function $F_{m}$ at the
$m^{th}$ RTO is a mapping from the space of measurements and received messages
to that of the reconstructed sequence denoted as%
\begin{equation}
F_{m}:\mathcal{Y}_{m}^{n}\times\mathcal{J}_{m}^{1}\times\ldots\times
\mathcal{J}_{m}^{K}\rightarrow\mathcal{\hat{X}}_{m}^{n},\text{ \ }m=1,2.
\label{Dec}%
\end{equation}
Let $J_{m}\equiv\left\{  J_{m}^{k}\right\}  _{k=1}^{M}$, denote the set of all
indices communicated by the $m^{th}$ RTO, $m=1,2$, such that $M_{1}$ and
$M_{2}$ denote the size of $J_{1}$ and $J_{2}$, respectively. The expected
distortion $D\,_{m}\ $at the $m^{th}$ RTO is given by
\begin{equation}
D_{m}=\frac{1}{n}\mathbb{E}\left[
{\textstyle\sum\limits_{i=1}^{n}}
\left(  X_{m,i}-\hat{X}_{m,i}\right)  ^{2}\right]  \text{, }m=1,2,
\label{Dist}%
\end{equation}
and the privacy leakage$,$ $L_{1}$, about state $1$ at RTO $2$ is given by%

\begin{equation}
L_{1}=\frac{1}{n}I\left(  X_{1}^{n};J_{1},J_{2},Y_{2}^{n}\right)  ,
\end{equation}
and the privacy leakage$,$ $L_{2}$, about state $2$ at RTO $1$ is given by%
\begin{equation}
L_{2}=\frac{1}{n}I\left(  X_{2}^{n};J_{1},J_{2},Y_{1}^{n}\right)  . \label{Eq}%
\end{equation}
The communication rate of the $m^{th}$ RTO is denoted by
\begin{equation}
R_{m}=n^{-1}\log_{2}M_{m},m=1,2. \label{CommRate}%
\end{equation}

\begin{definition}
\label{DefUP}The utility-privacy tradeoff region $\mathcal{T}$ is the set of
all $\left(  D_{1},D_{2},L_{1},L_{2}\right)  $ for which there exists a coding
scheme given by (\ref{Enc1})-(\ref{Dec}) with parameters $(n,K,M_{1}%
,M_{2},D_{1}+\epsilon,D_{2}+\epsilon,L_{1}+\epsilon,L_{2}+\epsilon)$ for $n$
sufficiently large such that $\epsilon\rightarrow0$ as $n\rightarrow\infty$.
\end{definition}

\section{\label{SecMR}Main Results}

The coding and communication protocol described in the previous section is
constrained only by the fidelity requirements of the desired state estimate at
each RTO and the privacy leakage of the state at the other RTO. Furthermore,
from Definition \ref{DefUP}, we note that the utility-privacy tradeoff region
$\mathcal{T}$ is also dependent on size of the encoding indices $M_{1}$ and
$M_{2}\,,$ i.e., for every $\left(  D_{1},D_{2},L_{1},L_{2}\right)
\in\mathcal{T}$, there can be several achievable coding schemes each
corresponding to some communication rate pair $\left(  R_{1},R_{2}\right)  $.
Thus, it suffices to determine the set of all rate-distortion-leakage $\left(
R_{1},R_{2},D_{1},D_{2},L_{1},L_{2}\right)  $ tuples to determine the
utility-privacy tradeoff region.

\subsection{Rate-Distortion-Leakage (RDL) Tradeoff}

Recall that $Y_{1}=X_{1}+\alpha X_{2}+N_{1}$ and $Y_{2}=\beta X_{1}%
+X_{2}+N_{2}$, where $X_{m}\sim\mathcal{N}(0,1),m=1,2$, and $N_{m}%
\sim\mathcal{N}(0,\sigma_{m}^{2}),m=1,2$, are all mutually independent. We
first state the range of allowable distortion pairs $\left(  D_{1}%
,D_{2}\right)  $ as%
\[
D_{m}\in\lbrack D_{\min,m},D_{\max,m}],m=1,2,
\]
where
\begin{subequations}
\label{Dmins}%
\begin{align}
D_{\min,1} &  =1-\frac{(\beta^{2}V_{1}+V_{2}-2\beta E)}{(V_{1}V_{2}-E^{2}%
)}\label{Dmin1}\\
D_{\min,2} &  =1-\frac{(V_{1}+\alpha^{2}V_{2}-2\alpha E)}{(V_{1}V_{2}-E^{2}%
)},\label{Dmin2}%
\end{align}
and
\end{subequations}
\begin{subequations}
\begin{equation}
D_{\max,1}=1-\frac{1}{V_{1}},\quad D_{\max,2}=1-\frac{1}{V_{2}},\label{Dmaxs}%
\end{equation}
for
\end{subequations}
\begin{subequations}
\label{Vdef}%
\begin{align}
V_{1} &  \equiv1+\alpha^{2}+\sigma_{1}^{2},\\
V_{2} &  \equiv1+\beta^{2}+\sigma_{2}^{2}\text{, and}\\
E &  \equiv\alpha+\beta.
\end{align}
The lower bounds $D_{\min,1}$ and $D_{\min,2}$ on $D_{1}$ and $D_{2}$ in
(\ref{Dmins}), respectively, are obtained by considering an enhanced system,
in which both observations $(Y_{1},Y_{2})$ can be jointly used for estimating
$X_{1}$ and $X_{2}$. For such a system, the minimum mean square error
(MMSE)\ estimator minimizes the quadratic distortion in (\ref{Dist}).
Therefore, $D_{\min,1}$ and $D_{\min,2}$ are given by%
\end{subequations}
\begin{align}
D_{\min,1} &  =\mathbb{E}\left[  var(X_{1}|Y_{1},Y_{2})\right]  \\
D_{\min,2} &  =\mathbb{E}\left[  var(X_{2}|Y_{1},Y_{2})\right]
\end{align}
where for two random variables $A$ and $B$, $var\left(  A|B\right)  $ is the
conditional variance of $A$ conditioned on $B.$ On the other hand, $D_{\max
,1}$ and $D_{\max,2}$ correspond to the fidelity criterion achievable using
only the locally available measurements at each RTO such that%
\begin{align}
D_{\max,1} &  =\mathbb{E}\left[  var(X_{1}|Y_{1})\right]  \\
D_{\max,2} &  =\mathbb{E}\left[  var(X_{2}|Y_{2})\right]  .
\end{align}

The range of feasible leakage values $L_{1}$ and $L_{2}$ are
\begin{align}
L_{1}  &  \in\lbrack I(X_{1};Y_{2}),I(X_{1};Y_{1},Y_{2})],\text{ and}\\
L_{2}  &  \in\lbrack I(X_{2};Y_{1}),I(X_{2};Y_{1},Y_{2})].
\end{align}
The lower bound on $L_{1}$, i.e., $I(X_{1};Y_{2})$ is guaranteed \emph{only}
for the case in which there is no public communication. This scenario can
arise if $D_{1}$ and $D_{2}$ can be achieved at each RTO with its own
measurements (corresponding to $D_{\max,1}$ and $D_{\max,2}$, respectively).
At the other extreme, the maximum value of leakage $L_{1}$ can go up to
$I(X_{1};Y_{1},Y_{2})$. This corresponds to an infinite rate transmission from
RTO $1$ to RTO $2$ to satisfy the distortion requirement $D_{2}=D_{\min,2}$. A
similar upper bounding argument holds for $L_{2}$.

Finally, we exploit the Gaussian nature of the system variables to write
\begin{subequations}
\label{Xfunc}%
\begin{align}
E[X_{1}|Y_{1},Y_{2}]  &  =k_{1}Y_{1}+k_{2}Y_{2}\\
E[X_{2}|Y_{1},Y_{2}]  &  =l_{1}Y_{1}+l_{2}Y_{2}%
\end{align}
where
\end{subequations}
\begin{subequations}
\label{kandl}%
\begin{align}
k_{1}  &  =\frac{V_{2}-\beta E}{V_{1}V_{2}-E^{2}},\quad k_{2}=\frac{\beta
V_{1}-E}{V_{1}V_{2}-E^{2}},\hbox{ and}\\
l_{1}  &  =\frac{\alpha V_{2}-E}{V_{1}V_{2}-E^{2}},\quad l_{2}=\frac
{V_{1}-\alpha E}{V_{1}V_{2}-E^{2}}.
\end{align}
Using these results, we now summarize the rate-distortion-leakage tradeoff for
the problem under consideration in the following theorem.
\end{subequations}
\begin{theorem}
\label{theorem1} The rate-distortion-leakage tradeoff $\left(  R_{1}%
,R_{2},D_{1},D_{2},L_{1},L_{2}\right)  $ is given as follows.

If $D_{2}\in\lbrack D_{\min,2},D_{\max,2}]$, then
\begin{align}
R_{1}  &  =\frac{1}{2}\log\left(  \frac{(V_{1}V_{2}-E^{2})l_{1}^{2}}%
{V_{2}(D_{2}-D_{\min,2})}\right)  ,\text{ and}\label{Th_R1}\\
L_{1}  &  =\frac{1}{2}\log\left(  \frac{l_{1}^{2}}{l_{1}^{2}D_{\min,1}%
+k_{1}^{2}(D_{2}-D_{\min,2})}\right)  . \label{Th_L1}%
\end{align}

If $D_{2}\geq D_{\max,2}$, then
\begin{align}
R_{1}  &  =0,\text{ and}\\
L_{1}  &  =\frac{1}{2}\log\left(  V_{2}\left/  \left(  V_{2}-\beta\right)
\right.  \right)  .
\end{align}

If $D_{1}\in\lbrack D_{\min,1},D_{\max,1}]$,
\begin{align}
R_{2}  &  =\frac{1}{2}\log\left(  \frac{(V_{1}V_{2}-E^{2})k_{2}^{2}}%
{V_{1}(D_{1}-D_{\min,1})}\right)  ,\text{ and}\label{Th_R2}\\
L_{2}  &  =\frac{1}{2}\log\left(  \frac{k_{2}^{2}}{k_{2}^{2}D_{\min,2}%
+l_{2}^{2}(D_{1}-D_{\min,1})}\right)  . \label{Th_L2}%
\end{align}

If $D_{1}\geq D_{\max,1}$, then
\begin{align}
R_{2}  &  =0,\text{ and}\\
L_{2}  &  =\frac{1}{2}\log\left(  V_{1}\left/  \left(  V_{1}-\alpha\right)
\right.  \right)  ,
\end{align}
where $D_{\min,m},$ $D_{\max,m}$, $V_{m}$, $E,$ and $l_{m}$, $k_{m}$, $m=1,2,$
are defined in (\ref{Dmins}), (\ref{Dmaxs}), (\ref{Vdef}), and (\ref{kandl}).
\end{theorem}

\begin{remark}
Theorem \ref{theorem1} shows that the optimal rate-distortion-leakage tradeoff
for the Gaussian case admits a decoupling with respect to the distortion pair
$(D_{1},D_{2})$. By decoupling, we mean that the leakage of source $X_{1}$ and
the transmission rate of RTO $1$ depends \emph{solely} on the distortion
$D_{2}$ desired by RTO $2$. Similarly, the leakage of $X_{2}$ and the
transmission rate of RTO $2$ depends \emph{solely} on the distortion $D_{1}$
desired by RTO $1$. The reason for this decoupling can be attributed to the no
rate-loss property of jointly Gaussian sources, which can be described as
follows: if RTO $2$ is interested in reconstructing $X_{2}$ at a distortion
$D_{2}$, then the minimal rate of transmission required by RTO $1$ remains
unchanged even if the source $Y_{2}$ is available at RTO $1$.
\end{remark}

We briefly sketch the proof below; we first derive lower bounds on the rate
and privacy leakage and present a coding scheme that achieves them.

\begin{proof}
\textit{Lower bound on }$R_{1},R_{2}:$ Recall that we write $J_{1}$ to denote
the message from RTO 1 over all $K$ communication rounds. Given a distortion
pair $(D_{1},D_{2})$, we prove the lower bound on $R_{1}$ as follows:
\begin{subequations}
\label{R1bound}%
\begin{align}
nR_{1}  &  =H(J_{1})\\
&  \geq I(Y_{1}^{n};J_{1}|Y_{2}^{n})\\
&  \geq nh(Y_{1}|Y_{2})-\sum_{t=1}^{n}h(Y_{1t}|J_{1},Y_{2}^{n})\\
&  \geq nh(Y_{1}|Y_{2})-\sum_{t=1}^{n}\frac{1}{2}\log\left(  2\pi
e\mbox{Var}(Y_{1t}|J_{1},Y_{2}^{n})\right) \\
&  \geq nh(Y_{1}|Y_{2})-\frac{n}{2}\log\left(  2\pi e\sum_{t=1}^{n}%
\mbox{Var}(Y_{1t}|J_{1},Y_{2}^{n})/n\right) \\
&  \geq nh(Y_{1}|Y_{2})-\frac{n}{2}\log\left(  2\pi e(D_{2}-D_{\min,2}%
)/l_{1}^{2}\right) \\
&  =\frac{n}{2}\log(2\pi e(V_{1}V_{2}-E^{2})/V_{2})\nonumber\\
&  \quad-\frac{n}{2}\log\left(  2\pi e(D_{2}-D_{\min,2})/l_{1}^{2}\right) \\
&  =\frac{n}{2}\log\left(  \frac{(V_{1}V_{2}-E^{2})l_{1}^{2}}{V_{2}%
(D_{2}-D_{\min,2})}\right)
\end{align}
where the inequalities in (\ref{R1bound}) follow from the chain rule, the fact
that conditioning does not increase entropy, the fact that Gaussian
distribution maximizes the differential entropy for a given variance, and the
concavity of the $\log$ function. The variance of $Y_{1t}$ conditioned on
$J_{1}$ and $Y_{2}^{n}$ can be computed using (\ref{Xfunc}) and (\ref{kandl})
and is omitted due to space limitations. A lower bound on $R_{2}$ can be
obtained similarly. Therefore, $R_{1}$ and $R_{2}$ can be lower bounded by the
expressions in (\ref{Th_R1}) and (\ref{Th_R2}), respectively.

\textit{Lower bounds on }$\left(  L_{1},L_{2}\right)  :$ Given an arbitrary
code that achieves a certain distortion pair $(D_{1},D_{2})$, we derive lower
bound on the leakages for both RTOs as follows:
\end{subequations}
\begin{subequations}
\label{lBOUND}%
\begin{align}
L_{1}  &  \geq\frac{1}{n}I(X_{1}^{n};J_{1},J_{2},Y_{2}^{n})\\
&  \geq h(X_{1})-\frac{1}{n}h(X_{1}^{n}|J_{1},J_{2},Y_{2}^{n})\\
&  \geq h(X_{1})-\frac{1}{n}h(X_{1}^{n}|J_{1},Y_{2}^{n})\\
&  \geq h(X_{1})-\frac{1}{n}h(X_{1}^{n}|J_{1},Y_{2}^{n})\\
&  \geq h(X_{1})-\frac{1}{n}\sum_{t=1}^{n}h(X_{1t}|J_{1},Y_{2}^{n})\\
&  \geq h(X_{1})-\frac{1}{n}\sum_{t=1}^{n}\frac{1}{2}\log(2\pi
e\mbox{ Var}(X_{1t}|J_{1},Y_{2}^{n}))\\
&  \geq h(X_{1})-\frac{1}{2}\log\left(  2\pi e\frac{1}{n}\sum_{t=1}%
^{n}\mbox{ Var}(X_{1t}|J_{1},Y_{2}^{n})\right) \\
&  \geq h(X_{1})-\frac{1}{2}\log\left(  2\pi e(D_{\min,1}+k_{1}^{2}%
(D_{2}-D_{\min,2})/l_{1}^{2})\right)
\end{align}
where the inequalities in (\ref{lBOUND}) follow from the chain rule, the fact
that conditioning does not increase entropy, the fact that Gaussian
distribution maximizes the differential entropy for a given variance, and the
concavity of the $\log$ function. A lower bound on $L_{2}$ can be obtained
similarly. Simplifying further, we can lower bound $L_{1}$ and $L_{2}$ with
the expressions in (\ref{Th_L1}) and (\ref{Th_L2}), respectively.

\textit{Upper bounds on }$\left(  R\,_{1},R_{2}\right)  $ and $\left(
L_{1},L_{2}\right)  $ \textit{via an achievable coding scheme}: The lower
bounds derived above can be achieved using the Wyner-Ziv source coding scheme
with decoder side information \cite{Wyner_Ziv} at each RTO. We briefly
describe the encoding scheme at RTO 1 and the resulting decoding at RTO 2; the
coding scheme for RTO 2 follows analogously. The encoding is such that RTO $1$
generates a set of $M_{1}$ sequences $U_{1}^{n}\left(  j_{1}\right)  $,
$j_{1}=1,2,\ldots,M_{1}$, where $M_{1}=2^{n\left(  I(U_{1};Y_{1}%
)+\epsilon\right)  }$. However, to exploit the fact that RTO 2 has correlated
measurements, RTO $1$ further bins its $M_{1}$ sequences into $S_{1}$ bins
chosen at random where $S_{1}=2^{n\left(  I(U_{1};Y_{1})-I(U_{1}%
;Y_{2})+\epsilon\right)  }$. Upon observing a measurement sequence $y_{1}%
^{n},$ the encoder at\ RTO 1 searches for a $U_{1}^{n}\left(  j_{1}\right)  $
sequence such that $y_{1}^{n}$ and $u_{1}^{n}\left(  j_{1}\right)  $ are
jointly typical (as defined in \cite[Chap. 10]{CTbook2}) where the choice of
$M_{1}$ ensures that there exists at least one such $j_{1}$. Using the fact
that RTO 2 has its own measurements $Y_{2}^{n}$ as side information, the
encoder at RTO 1 sends only $b\left(  j_{1}\right)  $ where $b\left(
j_{1}\right)  $ is the bin index of the $u_{1}^{n}\left(  j_{1}\right)  $ sequence.

For Gaussian distributed state and measurements, the sequence $U_{1}^{n}$ is
chosen such that the `test channel' from $U_{1}$ to $Y_{1}$ yields
$Y_{1}=U_{1}+Q_{1}$, where $Q_{1}$ is a Gaussian random variable, independent
of $U_{1}$, with variance $S_{1}$ chosen to satisfy (\ref{Dist}) and
\end{subequations}
\begin{equation}
\hat{X}_{1}=\mathbb{E}\left[  X_{1}|U_{2},Y_{1}\right]  . \label{XhatGauss}%
\end{equation}
Using (\ref{Dist}) and (\ref{XhatGauss}), we see that the achievable
distortion $D_{1}$ is simply the conditional variance of $X_{1}$ given
$(Y_{1},U_{2})$. From (\ref{Dmin1}), since $D_{\min,1}$ is the conditional
variance of $X_{1}$ given $(Y_{1},Y_{2})$, we have that $D_{1}$ is given by
the expression in (\ref{Dmin1}) with $V_{2}$ replaced by $V_{2}+S_{2}$; thus,
the variance of the quantization noise $S_{2}$ is a function of $D_{1}$. One
can similarly obtain $D_{2}$ from (\ref{Dmin2}) by replacing $V_{1}$ by
$V_{1}+S_{1}$ and observe that $S_{1}$ depends on $D_{2}$ but not $D_{1}$. The
resulting rate $R_{1}$ is simply the Wyner-Ziv rate $R_{WZ}$ between encoder 1
and decoder 2 with side information $Y_{2}$ and is given by%
\begin{align}
R_{1}  &  =R_{WZ}=I(Y_{1};U_{1}|Y_{2})\\
&  =\frac{1}{2}\log\left(  \frac{var\left(  Y_{1}|Y_{2}\right)  }{var\left(
Y_{1}|U_{1}Y_{2}\right)  }\right)  .
\end{align}
Writing $S_{2}$ in terms of $D_{2}$ and $D_{\min,2}$ and simplifying the
expressions, one can verify that $R_{1}$ simplifies to the expression in
Theorem (\ref{Th_R1}). Since $U_{1}^{n}$ is a function of $\left(  J_{1}%
,Y_{2}^{n}\right)  $, the expression for $L_{1}$ defined in (\ref{Eq})
simplifies to $I\left(  X_{1};U_{1}Y_{2}\right)  $ which in turn further
simplifies to the expression in (\ref{Th_L1}). Finally, one can similarly show
that Wyner-Ziv encoding is rate and leakage optimal for RTO 2.
\end{proof}

\begin{remark}
The leakage of $X_{1}$ at RTO $2$ depends only on the desired distortion of
$X_{2}$ and vice-versa, or alternatively, the feasibility of the distortion
desired by RTO $2$ depends on the leakage permissible by RTO $1$. Thus,
cooperation to achieve a desired fidelity at the other RTO inevitably results
in a proportional privacy leakage.
\end{remark}

\begin{remark}
Practical implementations of the Wyner-Ziv coding schemes have been well
studied and can be applied to the problem at hand.
\end{remark}

\subsection{Illustration}

\begin{figure}
[ptb]
\begin{center}
\includegraphics[
trim=0.869929in 0.228697in 0.678595in 0.330604in,
height=2.3791in,
width=3.0372in
]%
{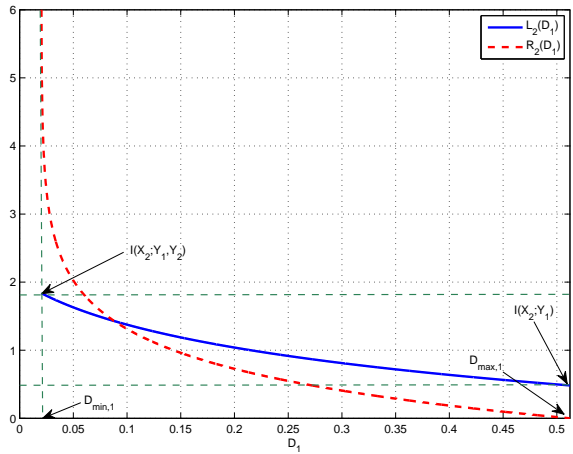}%
\caption{Plots of $R_{2}$ and $L_{2}$ vs. $D_{2}$.}%
\label{FigPlot1}%
\end{center}
\end{figure}

\begin{figure}
[ptb]
\begin{center}
\includegraphics[
trim=0.879283in 0.231660in 0.705807in 0.353118in,
height=2.341in,
width=3.0217in
]%
{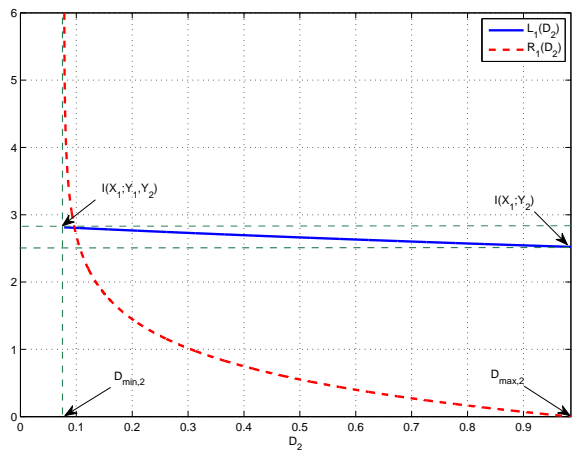}%
\caption{Plots of $R_{1}$ and $L_{1}$ vs. $D_{2}$.}%
\label{FigPlot2}%
\end{center}
\end{figure}

We illustrate our results for $\alpha=1,$ $\beta=8,$ $\sigma_{1}^{2}=0.05,$
and $\sigma_{2}^{2}=1.$ The parameters so chosen demonstrate the need for
communication between the RTOs since at each RTO, there is interference from
the measurements at the other RTO as well as noise. In Fig. \ \ref{FigPlot1},
$R_{1}$ and $L_{1}$ are plotted as functions of $D_{2}$ while in Fig.
\ref{FigPlot2}, the effect of changing the distortion $D_{1}$ from its minimum
to its maximum value on the rate $R_{2}$ and privacy leakage $L_{2}$ for RTO 2
are shown. The asymmetry in the interference and noise levels is captured in
the two plots.%

%

\section{\label{SecCR}Concluding Remarks}

In this paper, we have formalized the conflicting objectives of estimation
accuracy and competitive privacy in smart-grid operations in terms of a class
of information theoretic multi-terminal source coding problems. For a scalar
Gaussian source model with two RTOs, we have completely characterized the set
of optimal privacy-utility tuples via the rate-distortion-leakage tradeoff for
a mean-squared estimation criterion. We have shown that the RDL tradeoff at
each RTO\ depends only on the fidelity requirement at the other RTO with a
higher fidelity requirement leading to a higher rate and higher leakage and
vice-versa. While the results here also extend to vector Gaussian sources and
a mean-square fidelity criterion, we intend to investigate the RDL trade-off
for general non-Gaussian source models and multiple RTOs in a forthcoming paper.

Finally, we comment on some practical implementation issues associated with
the proposed distributed formulation. While the optimal encoding strategies
are necessarily block based, in practice it may not be feasible to accumulate
the incoming data for long at each RTO and subsequently encode the entire
block. A potential research direction would be the investigation of heuristic
real-time encoding schemes that yield close to optimal performance with much
less encoding complexity. Finally, a practical approach for the distributed
and decentralized information broadcast in such multi-agent networks is via
gossip protocols (see, for example, \cite{Boyd-GossipInfTheory}).

\bibliographystyle{IEEEtran}
\bibliography{refsSG}

\end{document}